\documentclass[journal,12pt,draftclsnofoot,onecolumn]{IEEEtran}

\usepackage{amsmath}
\usepackage{amssymb}
\usepackage{mathrsfs}
\usepackage{cite}
\usepackage{epsfig}
\usepackage{epsfig}
\usepackage{graphics}
\usepackage{hyperref}
\usepackage{epsfig}
\usepackage{setspace}

\newtheorem{lem}{Lemma}

\newtheorem{defi}{Definition}

\begin{document}

\title{Optimum Relay Scheme in a Secure Two-Hop Amplify and Forward Cooperative Communication System}
\author{Ghadamali Bagherikaram, Konstantinos N. Plataniotis\\
The Edward S. Rogers Sr. Department of ECE, University of Toronto,\\
 10 King's College Road, Toronto, Ontario, Canada M5S 3G4\\
 Emails: \{gbagheri, kostas\}@comm.utoronto.ca
}
 \maketitle

\begin{abstract}
A MIMO secure two-hop wireless communication system is considered in this paper. In this model, there are no direct links between the source-destination and the source-eavesdropper. The problem is maximizing the secrecy capacity of the system over all possible amplify and forward (AF) relay strategies, such that the power consumption at the source node and the relay node is limited. When all the nodes are equipped with single antenna, this non-convex optimization problem is fully characterized. When all the nodes (except the intended receiver) are equipped with multiple antennas, the optimization problem is characterized based on the generalized eigenvalues-eigenvectors of the channel gain matrices.
\end{abstract}

\section{Introduction}

The use of relay nodes in communication systems has offered
significant performance benefits, including being able to
achieve spatial diversity through node cooperation \cite{1,2}
and extending coverage without requiring large transmitter
powers. Cooperative communication, therefore, has been an attractive option for use in
cellular, ad-hoc networks and military communication systems \cite{3}.
The most common relaying protocols are Decode and
Forward (DF) and Amplify and Forward (AF). The AF scheme is a
simple scheme, which amplifies the signal transmitted from the
source and forwards it to the destination \cite{4,5}, and unlike the DF
scheme, no decoding at the relay is performed. AF techniques
may use the knowledge of the statistics of the noise in addition to the knowledge of all channel state information to assist in the amplification of the signal.

The notion of information theoretic secrecy in communication systems was first introduced in \cite{6}. The information theoretic secrecy requires that the received signal
by an eavesdropper not provide any information about the transmitted messages.
Following the pioneering works of \cite{7} and \cite{8} which have studied the wiretap
channel, many extensions of the wiretap channel model have been considered from a
perfect secrecy point of view (see e.g., \cite{9,10}).

Recently, AF MIMO relay systems have gained more attention from both academic and industrial communities, due to its simplicity and its benefits. The AF MIMO relay systems are, for example, adopted in future communication protocols such as LTE and IMT-advanced to enhance the coverage of base stations.

Most research works in MIMO cooperative communication systems have focused on the role of MIMO in enhancing the throughput and robustness. In this work, however, we focus on the role of such multiple antennas in enhancing wireless security. Particularly, in this paper, we consider a two-hop AF MIMO cooperative communication system in which there are no direct links between the source-destination and the source-eavesdropper nodes. Our goal is to maximize the physical layer security of the system with the constraint of limited available power at the source and the relay node. We formulate this problem as a non-convex constraint optimization problem. In the simplest scheme where all nodes have single antenna, we fully characterize the optimum relay strategy and show that when the available power is larger than a threshold, then the relay node does not need to consume all the available power. This is due to the fact that the secrecy capacity of the system will be saturated at high SNRs; therefore, using more power could not improve the  secrecy capacity of the system. We then consider a scenario in which all the nodes except the intended receiver are equipped with multiple antennas. The significance of this model is when a base station wishes to broadcast secure information by the help of a relay node to small mobile units. In this scenario small mobile units have single antenna while the base station, the relay node, and the eavesdropper can afford multiple antennas. We characterize the optimum relay matrix and the transmitter covariance matrix by using the generalized eigenvectors of the channel state information matrices. Finally, we explicitly illustrate the optimum power allocation at the relay node.

\section{Preliminaries and Related Works}

\subsection{Notation}
In this paper, a vector will be written as a small bold letter (e.g. $\mathbf{x}$)
and a matrix will be denoted by a capital bold letter (e.g. $\mathbf{A}$). The function $E[X]$ represents the statistical expectation of the random variable $X$ and the function $\hbox{Tr}[\mathbf{A}]$ represents the trace value of the matrix $\mathbf{A}$. For a given matrix $\mathbf{A}$, its determinant is represented by $|\mathbf{A}|$.

\subsection{System Model and Problem Statement}
In this paper, we consider a secure wireless communication system where a source is communicating with the destination through a relay within the presence of an eavesdropper, as shown in Fig.1.  In this model, we assume that all the nodes have $N$ antennas and there are no direct links between the source-destination and the source-eavesdropper. Such a system may occur in many practical
situations, as for example when two base stations equipped
with multiple antennas communicate with each other via a relay node in the presence of an illegitimate eavesdropper.
\begin{figure}
\leftline{\includegraphics[scale=.145]{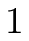}} \caption{Secure Two-Hop Amplify $\&$ Forward (AF) Network System Model.}
\end{figure}

The signal received by the destination and the eavesdropper are given as follows:
\begin{IEEEeqnarray}{lr}\label{model}
\mathbf{y}=\mathbf{H_{r}}\mathbf{G}(\mathbf{H_{1}}\mathbf{x}+\mathbf{n_{1}})+\mathbf{n_{r}}\\
\nonumber \mathbf{z}=\mathbf{H_{e}}\mathbf{G}(\mathbf{H_{1}}\mathbf{x}+\mathbf{n_{1}})+\mathbf{n_{e}}
\end{IEEEeqnarray}
where $\mathbf{x}$ is the transmitted $n\times 1$ vector, and $\mathbf{H_{1}}$, $\mathbf{H_{r}}$, $\mathbf{H_{e}}$ are $n\times n$ channel gain matrixes from the source to the relay, and from the relay to the legitimate receiver and the eavesdropper, respectively. We assume that the relay node is in full duplex mode and amplifies the received signal by $\mathbf{G}$ and forwards it simultaneously (i.e. $\mathbf{v}=\mathbf{G}\mathbf{u}$). $\mathbf{n_{1}}$, $\mathbf{n_{r}}$, and $\mathbf{n_{e}}$ are the additive i.i.d Gaussian noise vectors with zero mean and covariance matrix $\mathbf{I}$ at the relay node, the intended receiver, and the eavesdropper, respectively. We assume that the transmitter is transmitting with power $p$ and the relay node has access to maximum power of $P$, i.e.,
\begin{IEEEeqnarray}{rl}
\hbox{Tr}\left[\mathbf{Q}\stackrel{\triangle}{=}E\left[\mathbf{xx^{\dag}}\right]\right]&= P \\ \nonumber
\hbox{Tr}\left[\mathbf{G\left(H_{1}QH_{1}^{\dag}+I\right)G^{\dag}}\right] &\leq P.
\end{IEEEeqnarray}
In this paper we are interested in designing the optimum AF strategy for the relay node. We assume that the relay node has access to the all channel matrix gains. The secrecy Capacity of this system is given by the following optimization problem:

\begin{IEEEeqnarray}{lr}\label{pr}
C_{s}=\max_{\mathbf{Q}\succeq 0, \mathbf{G}}
&\log\frac{\left|\mathbf{H_{r}G(H_{1}QH_{1}^{\dag}+I)G^{\dag}H_{r}^{\dag}+I}\right|}{\left|\mathbf{H_{r}GG^{\dag}H_{r}^{\dag}+I}\right|}\\ \nonumber&- \log\frac{\left|\mathbf{H_{e}G(H_{1}QH_{1}^{\dag}+I)G^{\dag}H_{e}^{\dag}+I}\right|}{\left|\mathbf{H_{e}GG^{\dag}H_{e}^{\dag}+I}\right|}
\end{IEEEeqnarray}
Such that:
\begin{IEEEeqnarray}{rl}\label{cons}
\hbox{Tr}\left[\mathbf{Q}\right]\leq P,          \hbox{Tr}\left[\mathbf{G\left(H_{1}QH_{1}^{\dag}+I\right)G^{\dag}}\right]\leq P
\end{IEEEeqnarray}

The secrecy level of the confidential message $W$ is measured at the eavesdropper in terms of equivocation rate which is given by
\begin{IEEEeqnarray}{lr}\nonumber
R_{e}=H(W|\mathbf{Z}).
\end{IEEEeqnarray}

The perfect secrecy capacity $C_{s}$ is the maximum amount of
information that can be sent to the legitimate receiver in a
reliable and confidential manner.
\subsection{Related Works}
For the Gaussian Multiple-Input Multiple-Output Multiple-Eavesdropper (MIMOME) channel (without relay node) and its extensions the works \cite{9,10,11} have proved that the secrecy capacity of the channel is given by

\begin{IEEEeqnarray}{rl}
C_{s}=\max_{\mathbf{Q}\succeq 0, \hbox{Tr}[\mathbf{Q}]\leq P}\frac{1}{2}\log\frac{\left|\mathbf{H_{r}QH_{r}^{\dag}+I}\right|}{\left|\mathbf{H_{e}QH_{e}^{\dag}+I}\right|}.
\end{IEEEeqnarray}
The above optimization problem involves solving a nonconvex problem. Usually nontrivial
techniques and strong inequalities are used to solve the
optimization problems of this type. In \cite{11}, the authors successfully characterized the capacity expression of the Gaussian Multiple-Input Single-Output Multiple-Eavesdropper (MISOME)
channel in terms of generalized eigenvalues. We summarize the results of \cite{11} here.

\begin{defi}[Generalized eigenvalues]
For a Hermitian matrix $\mathbf{A}$ and positive definite matrix $\mathbf{B}$,
$(\lambda,\mbox{\boldmath{$\psi$}})$ is referred to as a generalized eigenvalue-eigenvector pair
of $(\mathbf{A},\mathbf{B})$ if $(\lambda,\mbox{\boldmath{$\psi$}})$ satisfy
\begin{IEEEeqnarray}{rl}
\mathbf{A}\mbox{\boldmath{$\psi$}} = \lambda\mathbf{B}\mbox{\boldmath{$\psi$}}.
\end{IEEEeqnarray}
\end{defi}
Generalized eigenvalues-eigenvectors have the following property:

\begin{lem}[Variational Characterization]
The generalized eigenvectors of $(\mathbf{A},\mathbf{B})$ are the stationary point solution
to a particular Rayleigh quotient. Specifically, the largest
generalized eigenvalue is the maximum of the Rayleigh
quotient
\begin{IEEEeqnarray}{rl}
\lambda_{\max}(\mathbf{A},\mathbf{B}) = \max_{\mbox{\boldmath{$\psi$}}}\frac{\mbox{\boldmath{$\psi$}}^{\dag}\mathbf{A}\mbox{\boldmath{$\psi$}}}{\mbox{\boldmath{$\psi$}}^{\dag}\mathbf{B}\mbox{\boldmath{$\psi$}}}
\end{IEEEeqnarray}
and the optimum is attained by the eigenvector corresponding
to $\lambda_{\max}(\mathbf{A},\mathbf{B})$.
\end{lem}

For the MISOME channel of
\begin{IEEEeqnarray}{rl}
\mathbf{y}&=\mathbf{h_{r}^{\dag}x}+\mathbf{n_{r}}\\ \nonumber
\mathbf{z}&=\mathbf{H_{e}x}+\mathbf{n_{e}},
\end{IEEEeqnarray}
\cite{11} showed that the secrecy capacity is given by
\begin{IEEEeqnarray}{rl}\label{misome}
C_{s}&=\max_{\mathbf{Q}\succeq 0}\frac{1}{2}\log\frac{\left|P\mathbf{h_{r}^{\dag}Qh_{r}}+1\right|}{\left|P\mathbf{H_{e}QH_{e}^{\dag}+I}\right|}\\ \nonumber &=\frac{1}{2}\log\lambda_{\max}\left(\mathbf{I}+P\mathbf{h_{r}h_{r}^{\dag}},\mathbf{I}+P\mathbf{H_{e}^{\dag}H_{e}}\right).
\end{IEEEeqnarray}
The optimum covariance matrix $\mathbf{Q}$ is given by
\begin{IEEEeqnarray}{rl}\label{ten}
\mathbf{Q}=P\mbox{\boldmath{$\psi$}}_{\max}\mbox{\boldmath{$\psi$}}_{\max}^{\dag},
\end{IEEEeqnarray}
where $\mbox{\boldmath{$\psi$}}_{\max}$ is the normalized \footnote{In this paper we assume that the generalized eigenvectors are always normalized, i.e., $\mbox{\boldmath{$\psi$}}^{\dag}\mbox{\boldmath{$\psi$}}=1$.} eigenvector corresponding to the pencil $\left(\mathbf{I}+P\mathbf{h_{r}h_{r}^{\dag}},\mathbf{I}+P\mathbf{H_{e}^{\dag}H_{e}}\right)$
\section{Single Antenna Scheme}

In this section we consider the problem described in Fig.1, when all the nodes are equipped by only one antenna. By using the Lagrange Multiplier, the optimization problem of (\ref{pr}) and its constraints (\ref{cons}) can therefore be written as the following unconstraint problem:
\begin{IEEEeqnarray}{rl}
C_{s}=&\max_{g}\log\frac{\left|g^{2}h_{r}^{2}\left(Ph_{1}^{2}+1\right)+1\right|}{\left|g^{2}h_{r}^{2}+1\right|}\\ \nonumber &-\log\frac{\left|g^{2}h_{e}^{2}\left(Ph_{1}^{2}+1\right)+1\right|}{\left|g^{2}h_{e}^{2}+1\right|}+\lambda\left(g^{2}\left(Ph_{1}^{2}+1\right)-P\right).
\end{IEEEeqnarray}
The optimum value for $g$ (which is denoted by $g^{*}$) is such that $\frac{\partial C_{s}}{\partial g}=0$. Thus, after some mathematics we have
\begin{IEEEeqnarray}{rl}\label{op}
\frac{Ph_{1}^{2}\left(1-g^{*4}h_{r}^{2}h_{e}^{2}\left(ph_{1}^{2}+1\right)\right)}{M}+\lambda\left(Ph_{1}^{2}+1\right)=0
\end{IEEEeqnarray}
where,
\begin{IEEEeqnarray}{rl}
M=&\left(g^{*2}h_{r}^{2}\left(Ph_{1}^{2}+1\right)+1\right)\left(g^{*2}h_{e}^{2}\left(Ph_{1}^{2}+1\right)+1\right)\\ \nonumber &\times\left(g^{*2}h_{r}^{2}+1\right)\left(g^{*2}h_{e}^{2}+1\right).
\end{IEEEeqnarray}
Therefore, the optimum value for the relay gain is given as follows: When $\lambda \neq 0$, then the constraint must be satisfied; otherwise when $\lambda=0$, $g^{*}$ can be calculated from (\ref{op}). Hence,
\begin{IEEEeqnarray}{rl}
g^{*2}=\left\{
        \begin{array}{ll}
          \frac{1}{h_{r}h_{e}\sqrt{Ph_{1}^{2}+1}}, & \hbox{$\lambda=0$;} \\
          \frac{P}{Ph_{1}^{2}+1}, & \hbox{$\lambda \neq 0$.}
        \end{array}
      \right.
\end{IEEEeqnarray}
As always, $g^{*}$ must be such that $g^{*2}\leq \frac{P}{Ph_{1}^{2}+1}$, equivalently the above equation can be written as follows:
\begin{IEEEeqnarray}{rl}
g^{*2}=\left\{
        \begin{array}{ll}
          \frac{1}{h_{r}h_{e}\sqrt{Ph_{1}^{2}+1}}, & \hbox{$P\geq \frac{h_{1}^{2}+\sqrt{h_{1}^{4}+4h_{r}^{2}h_{e}^{2}}}{2h_{r}^{2}h_{e}^{2}}$;} \\
          \frac{P}{Ph_{1}^{2}+1}, & \hbox{$0\leq P \leq \frac{h_{1}^{2}+\sqrt{h_{1}^{4}+4h_{r}^{2}h_{e}^{2}}}{2h_{r}^{2}h_{e}^{2}}$.}
        \end{array}
      \right.
\end{IEEEeqnarray}

As we can see from the above equation, when the power $P$ is large enough, i.e., $P\geq\frac{h_{1}^{2}+\sqrt{h_{1}^{4}+4h_{r}^{2}h_{e}^{2}}}{2h_{r}^{2}h_{e}^{2}}$, the relay node does not need to use al the power $P$. This is due to the fact that the secrecy capacity will saturate for high $P$s.

\section{Multiple-Input Multiple-Relay Single-Output Multiple-Eavesdropper (MIMRSOME) Scheme }
In this section we consider the problem of Fig.1, when all the nodes except the intended receiver have access to multiple antenna. This model may occur in practice when the intended receiver is a mobile small unit that only can handle one antenna while the base station, the relay node, and the adversary eavesdropper are equipped by multiple antennas. The received signal by the legitimate receiver and the eavesdropper is the same as (\ref{model}) except that here, $\mathbf{H_{r}}$ is replaced by an $1\times n$ vector of $\mathbf{h_{r}^{\dag}}$ and the $n\times 1$ noise vector $\mathbf{n_{r}}$ is replaced by a scalar noise $n_{r}$.

The optimization problem of (\ref{pr}) and its constraints (\ref{cons}) can therefore be written as follows:
\begin{IEEEeqnarray}{rl}
C_{s}=\max_{\mathbf{Q}\succeq 0, \mathbf{G}}
&\log\frac{\left|\mathbf{h_{r}^{\dag}G\left(H_{1}QH_{1}^{\dag}+I\right)G^{\dag}h_{r}+I}\right|}{\left|\mathbf{H_{e}G(H_{1}QH_{1}^{\dag}+I)G^{\dag}H_{e}^{\dag}+I}\right|}\\ \nonumber+ &\log\frac{\left|\mathbf{H_{e}GG^{\dag}H_{e}^{\dag}+I}\right|}{\left|\mathbf{h_{r}^{\dag}GG^{\dag}h_{r}+I}\right|}
\end{IEEEeqnarray}
Such that:
\begin{IEEEeqnarray}{rl}\label{con2}
\hbox{Tr}\left[\mathbf{Q}\right]\leq P,          \hbox{Tr}\left[\mathbf{G\left(H_{1}QH_{1}^{\dag}+I\right)G^{\dag}}\right]\leq P.
\end{IEEEeqnarray}
Let us define $\mathbf{Q_{1}}$ and $\mathbf{Q_{2}}$ as follows:

\begin{IEEEeqnarray}{rl}
\mathbf{Q_{1}}&\stackrel{\triangle}{=}\mathbf{G\left(H_{1}QH_{1}^{\dag}+I\right)G^{\dag}}\\ \nonumber
\mathbf{Q_{2}}&\stackrel{\triangle}{=}\mathbf{GG^{\dag}}.
\end{IEEEeqnarray}

According to (\ref{con2}), $\mathbf{Q_{1}}$ and $\mathbf{Q_{2}}$ must be such that
\begin{IEEEeqnarray}{rl}
\hbox{\hbox{Tr}}[\mathbf{Q_{1}}]&\leq P,\\ \label{tr}
\hbox{\hbox{Tr}}\left[\mathbf{GH_{1}QH_{1}^{\dag}G^{\dag}}\right]+\hbox{Tr}\left[\mathbf{Q_{2}}\right]&\leq P.
\end{IEEEeqnarray}

Let us define $x$ as,
\begin{IEEEeqnarray}{rl}
x\stackrel{\triangle}{=}\hbox{Tr}\left[\mathbf{GH_{1}QH_{1}^{\dag}G^{\dag}}\right].
\end{IEEEeqnarray}

The optimization problem is therefore as follows:
\begin{IEEEeqnarray}{rl}
C_{s}=\max_{\mathbf{Q_{1}}\succeq 0, \mathbf{Q_{2}}\succeq 0}
&\log\frac{\left|\mathbf{h_{r}^{\dag}\mathbf{Q_{1}}h_{r}+I}\right|}{\left|\mathbf{H_{e}\mathbf{Q_{1}}H_{e}^{\dag}+I}\right|}\\ \nonumber+ &\log\frac{\left|\mathbf{H_{e}\mathbf{Q_{2}}H_{e}^{\dag}+I}\right|}{\left|\mathbf{h_{r}^{\dag}\mathbf{Q_{2}}h_{r}+I}\right|}
\end{IEEEeqnarray}
such that:
\begin{IEEEeqnarray}{rl}
\hbox{Tr}\left[\mathbf{Q_{1}}\right]\leq P,          \hbox{Tr}\left[\mathbf{\mathbf{Q_{2}}}\right]\leq P-x.
\end{IEEEeqnarray}

This problem can be viewed as two independent optimization problems of type problem (\ref{misome}). Let $(\lambda_{\max},\mbox{\boldmath{$\psi$}}_{\max})$ be the maximum generalized eigenvalue and its corresponding eigenvector of the following pencil:
\begin{IEEEeqnarray}{rl}
(\mathbf{I}+P\mathbf{h_{r}h_{r}^{\dag}},\mathbf{I}+P\mathbf{H_{e}^{\dag}H_{e}}).
\end{IEEEeqnarray}
Similarly, let $(\gamma_{\max},\mbox{\boldmath{$\varphi$}}_{\max})$ be the maximum generalized eigenvalue and its corresponding eigenvector of the following pencil:
\begin{IEEEeqnarray}{rl}\label{pencil}
(\mathbf{I}+(P-x)\mathbf{H_{e}^{\dag}H_{e}},\mathbf{I}+(P-x)\mathbf{h_{r}h_{r}^{\dag}}).
\end{IEEEeqnarray}

Therefore, according to (\ref{ten}) the optimum values for $\mathbf{Q_{1}}$ and $\mathbf{Q_{2}}$ are given by
\begin{IEEEeqnarray}{rl}\label{x}
\mathbf{Q_{1}}&=\mathbf{G\left(H_{1}QH_{1}^{\dag}+I\right)G^{\dag}}=P\mbox{\boldmath{$\psi$}}_{\max}\mbox{\boldmath{$\psi$}}_{\max}^{\dag}\\ \label{gg} \mathbf{Q_{2}}&=\mathbf{GG^{\dag}}=\left(P-x\right)\mbox{\boldmath{$\varphi$}}_{\max}\mbox{\boldmath{$\varphi$}}_{\max}^{\dag},
\end{IEEEeqnarray}
and the secrecy capacity is as follows:
\begin{IEEEeqnarray}{rl}
C_{s}=\frac{1}{2}\log\lambda_{\max}\gamma_{\max}.
\end{IEEEeqnarray}
A straightforward solution for the equation of (\ref{gg}) is as follows:
\begin{IEEEeqnarray}{rl}\label{g}
\mathbf{G}=\sqrt{P-x}\left[\mbox{\boldmath{$\varphi$}}_{\max}|~\mathbf{0}_{n\times(n-1)}\right],
\end{IEEEeqnarray}
where $\mathbf{0}_{n\times(n-1)}$ is an $n\times(n-1)$ all zero matrix. By substituting (\ref{g}) into (\ref{x}), the covariance matrix of $\mathbf{Q}$ is given by
\begin{IEEEeqnarray}{rl}
\mathbf{Q}=\left(\mathbf{GH_{1}}\right)^{-1}&\left[P\mbox{\boldmath{$\psi$}}_{\max}\mbox{\boldmath{$\psi$}}_{\max}^{\dag}-\left(P-x\right)\mbox{\boldmath{$\varphi$}}_{\max}\mbox{\boldmath{$\varphi$}}_{\max}^{\dag}\right]\\ \nonumber &\times\left(\mathbf{H_{1}^{\dag}G^{\dag}}\right)^{-1},
\end{IEEEeqnarray}
where $\mathbf{G}$ is as (\ref{g}).

To calculate the optimum value of $x$, note that
\begin{IEEEeqnarray}{rl}\label{ineq}
x&=\hbox{Tr}\left[\mathbf{GH_{1}QH_{1}^{\dag}G^{\dag}}\right]\\ \nonumber
&\stackrel{(a)}{=}\hbox{Tr}\left[\mathbf{H_{1}^{\dag}G^{\dag}GH_{1}Q}\right]\\ \nonumber
&\stackrel{(b)}{=}\hbox{Tr}\left[\left(P-x\right)\mathbf{H_{1}^{\dag}\mbox{\boldmath{$\psi$}}_{\max}\mbox{\boldmath{$\psi$}}_{\max}^{\dag}H_{1}Q}\right]\\ \nonumber&\stackrel{(c)}{\leq}\left(P-x\right)\left[\hbox{Tr}\left[\mathbf{H_{1}^{\dag}\mbox{\boldmath{$\psi$}}_{\max}\mbox{\boldmath{$\psi$}}_{\max}^{\dag}H_{1}}\right]\hbox{Tr}\left[\mathbf{Q}\right]\right]\\ \nonumber
&\stackrel{(d)}{=}\left(P-x\right)P\|\mathbf{H_{1}}^{\dag}\mbox{\boldmath{$\psi$}}_{\max}\|^{2},
\end{IEEEeqnarray}
where $(a)$ follows from the fact that $\hbox{Tr}\left[\mathbf{AB}\right]=\hbox{Tr}\left[\mathbf{BA}\right]$, $(b)$ follows from (\ref{gg}), $(c)$ follows from the fact that for any $\mathbf{A\succ 0}$ and $\mathbf{B}\succ 0$, $\hbox{Tr}\left[\mathbf{AB}\right]\leq\hbox{Tr}\left[A\right]\hbox{Tr}\left[B\right]$, and $(d)$ follows from the fact that $\hbox{Tr}\left[\mathbf{AA^{\dag}}\right]=\|\mathbf{A}\|^{2}$, where $\|.\|$ represents the Frobenius norm of a matrix.

From the inequality of (\ref{ineq}), $x$ is bounded as follows:
\begin{IEEEeqnarray}{rl}\label{cc}
x\leq \frac{P^{2}}{1+P\|\mathbf{H_{1}}^{\dag}\mbox{\boldmath{$\psi$}}_{\max}\|^{2}}.
\end{IEEEeqnarray}

Now, we must choose the parameter $x$ such that maximizes $C_{s}$ which is given by
\begin{IEEEeqnarray}{rl}
C_{s}&=\frac{1}{2}\log\lambda_{\max}\gamma_{\max}\\ \nonumber
&=\frac{1}{2}\log\lambda_{\max}\frac{1+(P-x)\|\mathbf{H_{e}}\mbox{\boldmath{$\varphi$}}_{\max}(x)\|^{2}}{1+(P-x)|\mathbf{h_{r}^{\dag}}\mbox{\boldmath{$\varphi$}}_{\max}(x)|^{2}}
\end{IEEEeqnarray}
with the constraint of (\ref{cc}). Note that in the above optimization problem $\lambda_{\max}$ is independent of $x$, however, $\mbox{\boldmath{$\varphi$}}_{\max}$ is a function of $x$, i.e.,
\begin{IEEEeqnarray}{rl}
\mbox{\boldmath{$\varphi$}}_{\max}(x)= \hbox{arg} \max_{\mbox{\boldmath{$\varphi$}}}\frac{1+(P-x)\|\mathbf{H_{e}}\mbox{\boldmath{$\varphi$}}\|^{2}}{1+(P-x)|\mathbf{h_{r}^{\dag}}\mbox{\boldmath{$\varphi$}}|^{2}}.
\end{IEEEeqnarray}
Thus, to abstain the optimum value of $x$ we need a more explicit relationship between $x$ and $\gamma_{\max}$. For this purpose, note that $(\gamma_{\max}, \mbox{\boldmath{$\varphi$}}_{\max})$ are the maximum generalized eigenvalue- eigenvector pair of pencil (\ref{pencil}). Hence,
\begin{IEEEeqnarray}{lr}\label{ee}
\left|\left(1-\gamma_{\max}\right)\mathbf{I}+\left(P-x\right)\left[\mathbf{H_{e}^{\dag}H_{e}}-\gamma_{\max}\mathbf{h_{r}h_{r}^{\dag}}\right]\right|\\ \nonumber =0.
\end{IEEEeqnarray}
In general we have,
\begin{IEEEeqnarray}{rl}
\left|\mathbf{I}+\mathbf{A}\right|=\sum_{k=0}^{\infty}\left(-\sum_{j=1}^{\infty}\frac{(-1)^{j}}{j!}\hbox{Tr}\left[\mathbf{A}^{j}\right]\right)^{k},
\end{IEEEeqnarray}
therefore equation (\ref{ee}) is equivalent to the following equation:
\begin{IEEEeqnarray}{lr}\nonumber
\sum_{k=0}^{\infty}\frac{1}{k!}\left(-\sum_{j=1}^{\infty}\frac{(-1)^{j}}{j!}\frac{P-x}{1-\gamma_{\max}}\hbox{Tr}\left[\left(\mathbf{H_{e}^{\dag}H_{e}}-\gamma_{\max}\mathbf{h_{r}h_{r}^{\dag}}\right)^{j}\right]\right)^{k}\\ \nonumber=0.
\end{IEEEeqnarray}

The above equation, explicitly, characterizes $\gamma_{\max}$ as a function of $x$ (or $P-x$). By solving this problem and setting $\frac{\partial\gamma_{\max}}{\partial x}=0$, we can determine the optimum power allocation of the relay node (which is $P-x$). Note that the optimum power allocation must always satisfy the constraint of (\ref{cc}).

\section{Conclusions}

We considered a two-hop AF MIMO cooperative communication system in which there are no direct links between the source-destination and the source-eavesdropper nodes. We maximized the physical layer security of the system with the constraint of limited available power at the source and the relay node. When all  the nodes have single antenna, we fully characterized the optimum relay strategy and showed that when the available power is larger than a threshold, then the relay node need not consume all the available power. Motivated by the fact that mobile units usually can afford only a single antenna, we then considered a scenario in which all the nodes (except the intended receiver) are equipped with multiple antennas. We characterized the optimum relay matrix and the transmitter covariance matrix by using the generalized eigenvectors of the channel state information matrices. Finally, we explicitly illustrated the optimum power allocation at the relay node.


\begin{thebibliography}{9}

\bibitem{1}
J. N. Laneman and G. W. Wornell, ``Exploiting distributed spatial
diversity in wireless networks," {\em in Proc. Allerton Conf. Communications,
Control, Computing,} Urbana-Champagne, IL, Sept. 2000.

\bibitem{2}
 A. Sendonaris, E. Erkip, and B. Aazhang, ``Increasing uplink capacity
via user cooperation diversity," {\em in Proc. IEEE Int. Symp. Inform. Theory
(ISIT),} Cambridge, MA, pp. 156, Aug. 1998.

\bibitem{3}
R. Pabst, B. H. Walke, D. C. Schultz, P. Herhold, H. Yanikomeroglu,
S. Mukherjee, H. Viswanathan, M. Lott, W. Zirwas, M. Dohler, H. Aghvami,
D. D. Falconer, and G. P. Fettweis, ``Relay-based deployment
concepts for wireless and mobile broadband radio," {\em IEEE Commun.
Mag,} vol. 42, no. 9, pp. 80-89, Sept. 2004.

\bibitem{4}
J. N. Laneman and G. W. Wornell, ``Energy efficient antenna sharing and
relaying for wireless networks," {\em in Proc. IEEE Wireless Communication
and Networking Conf. (WCNC),} Chicago, IL, Oct. 2000, pp. 7-12.

\bibitem{5}
J. N. Laneman, D. N. C. Tse, and G. W. Wornell, ``Cooperative diversity
in wireless neteworks efficient protocols and outage behavior," {\em IEEE
Trans. Inform. Theory,} vol. 50, no. 12, pp. 3062-3080, Dec. 2004.

\bibitem{6}
C. E. Shannon, ``Communication Theory of Secrecy Systems", {\em Bell
System Technical Journal}, vol. 28, pp. 656-715, October 1949.

\bibitem{7}
A. Wyner, ``The Wire-tap Channel", {\em Bell System Technical
Journal}, vol. 54, pp. 1355-1387, 1975

\bibitem{8}
I. Csisz´ar and J. K¨orner, ``Broadcast Channels with Confidential
Messages", {\em IEEE Trans. Inform. Theory}, vol. 24, no. 3, pp.
339-348, May 1978.

\bibitem{9}
A. Khisti, G. Wornell, A. Wiesel, and Y. Eldar, ``On the Gaussian MIMO Wiretap Channel", {\em in Proc. IEEE International Symposium on Information Theory} , pp.2471-2475, June 2007.

\bibitem{10}
G. Bagherikaram, A. S. Motahari and A. K. Khandani, ``The Secrecy Capacity Region of the Degraded Vector Gaussian Broadcast Channel", {\em in Proc. IEEE International Symposium on Information Theory}, pp.2772-2776, July 2009.

\bibitem{11}
A. Khisti and G. Wornell, ``Secure Transmission with Multiple Antennas: The MISOME Wiretap Channel", {\em EEE Trans. Inform. Theory}, Vol. 56, Issue.7, pp. 3088-3104, July 2010.

\end{thebibliography}
\end{document}